\documentclass[reviewcopy]{elsart}

\usepackage{graphicx}
\usepackage{enumerate}
\usepackage{graphicx}
\usepackage{amsmath}
\usepackage{amsfonts}
\usepackage{amssymb}
\newcommand{\ket}[1]{\left| #1 \right\rangle}
\newcommand{\bra}[1]{\left\langle #1 \right|}

\begin{document}
\begin{frontmatter}
\title{Nonadiabatic holonomic single-qubit gates in off-resonant $\Lambda$ systems}
\author[x]{Erik Sj\"oqvist}
\ead{erik.sjoqvist@kemi.uu.se}
\address[x]{Department of Physics and Astronomy, Uppsala University, Box 516, 
SE-751 20 Uppsala, Sweden}
\date{\today}
\begin{abstract}
We generalize nonadiabatic holonomic quantum computation in a resonant $\Lambda$ 
configuration proposed in [New J. Phys. 14 (2012) 103035] to the case of off-resonant 
driving lasers. We show that any single-qubit holonomic gate can be realized by 
separately varying the detuning, amplitude, and phase of the lasers. 
\end{abstract}
\end{frontmatter} 
\maketitle
Holonomic quantum computation (HQC) is the idea to use non-Abelian geometric phases to 
implement a universal set of quantum gates. It was first proposed in the context of adiabatic 
evolution by Zanardi and Rasetti \cite{zanardi99} based on Wilczek-Zee geometric phases 
\cite{wilczek84} associated with degenerate energy subspaces driven by slowly varying 
parameters. More recently, a scheme for fast holonomic quantum gates has been proposed 
\cite{sjoqvist12}. This scheme has subsequently been implemented experimentally 
\cite{abdumalikov13,feng13,arroyo-camejo14,zu14}. 

The nonadiabatic HQC scheme in Ref.~\cite{sjoqvist12} is based on a $\Lambda$ system 
driven by short resonant laser pulses. The high-speed feature makes the resulting gates 
potentially easier to implement as it implies a shorter exposure to detrimental 
decoherence effects. Here, we modify the original setup in Ref.~\cite{sjoqvist12} by allowing 
nonvanishing detuning of the two lasers driving off-resonant transitions between the excited 
state and the computational levels. This modified scheme can be used to implement
any single-qubit gate by separately varying the detuning, amplitude, and phase of the 
lasers, at the expense of restricting to square-shaped pulses. 

\begin{figure*}[ht]
\centering
\includegraphics[width=0.9\textwidth]{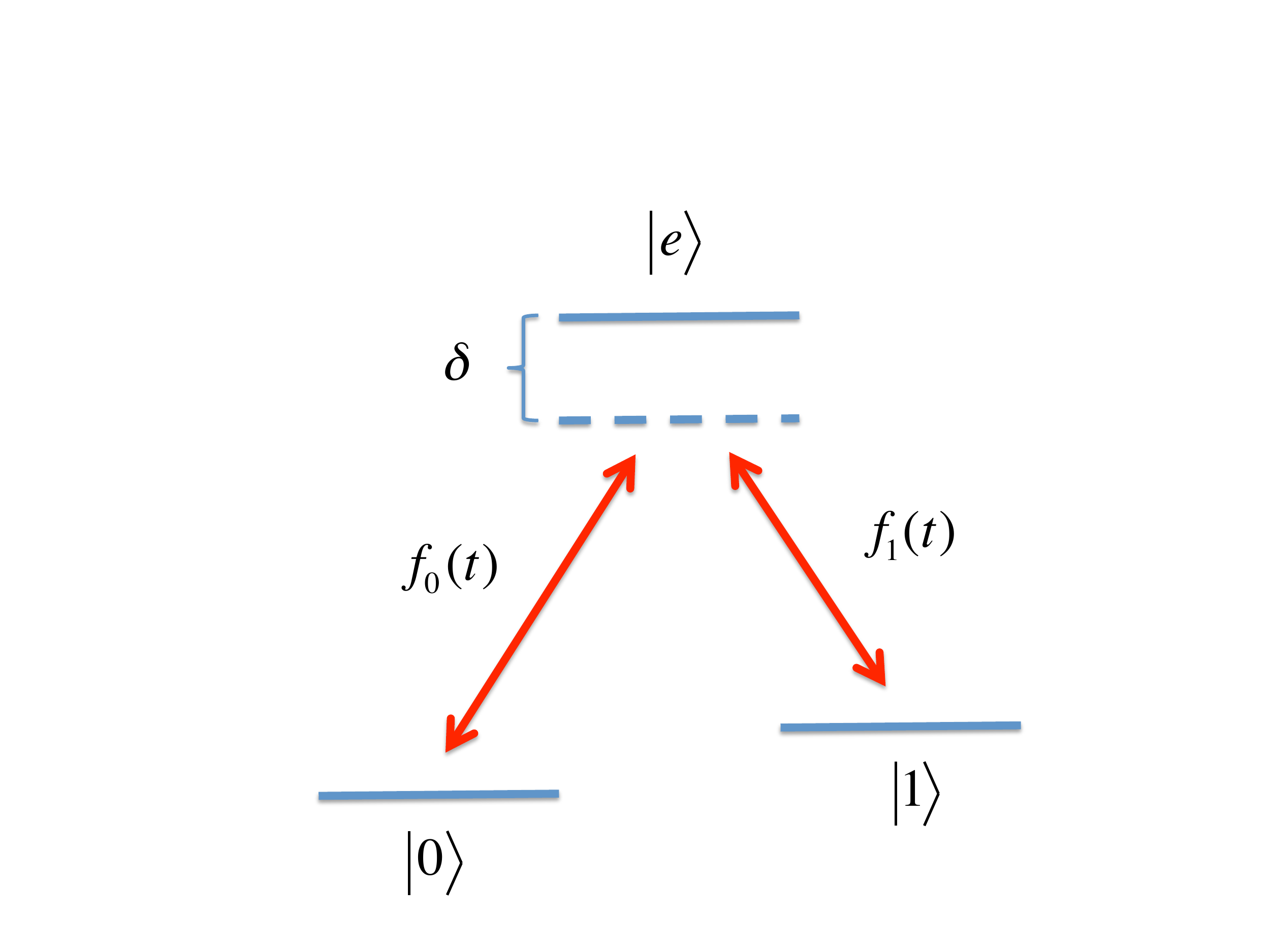}
\caption{Off-resonant $\Lambda$ system consisting of two energy levels $\ket{0}$ and 
$\ket{1}$ coupled to an excited state $\ket{e}$ by two pulsed laser beams with detuning 
$\delta$ and associated with Rabi frequencies $f_0 (t)$ and $f_1 (t)$.}
\label{fig:1}
\end{figure*}

Consider a quantum system exhibiting a three-level $\Lambda$-type configuration, in which 
two energy levels $\ket{0}$ and $\ket{1}$, spanning the computational state space, are coupled 
to an excited state $\ket{e}$ by two pulsed laser beams with detuning $\delta$ and associated 
with Rabi frequencies $f_0 (t)$ and $f_1 (t)$, see Fig.~\ref{fig:1}. The Hamiltonian in a frame that 
rotates with the laser fields reads ($\hbar = 1$ from now on)
\begin{eqnarray}
H(t) & = & F (t) \left( e^{-i\varphi} \sin \frac{\theta}{2} \ket{e} \bra{0} -
\cos \frac{\theta}{2} \ket{e} \bra{1} + \textrm{H.c.} \right) + \delta \ket{e} \bra{e} 
\nonumber \\
 & \equiv & F(t) H_0 + \delta \ket{e} \bra{e} ,
\label{eq:hamiltonian}
\end{eqnarray}
where rapidly oscillating counter-rotating terms have been neglected (rotating wave 
approximation) and we have put $f_0 (t) = F (t) e^{-i\varphi} \sin \frac{\theta}{2}$ and 
$f_1 (t) = -F (t) \cos \frac{\theta}{2}$. Here, $\theta$ and $\varphi$ are time independent 
over the duration of the pulse pair, which is controlled by the real-valued pulse envelope $F (t)$.

The standard form of nonadiabatic HQC in the $\Lambda$ configuration \cite{sjoqvist12} 
assumes that the lasers are on resonance with the transition frequencies, i.e., that the detuning 
$\delta$ vanishes. In this case, the evolution of the computational subspace $\mathcal{M} = 
\textrm{Span} \{ \ket{0},\ket{1} \}$ becomes purely geometric and cyclic with period $\tau$ 
such that $\int_0^{\tau} F (t) dt = \pi$, irrespective of the detailed form of $F (t)$. The path 
$C_{\bf n}$ traversed by the computational subspace in the Grassmannian 
$G(3;2)$\footnote{That is, the space of two-dimensional subspaces of 
a three-dimensional complex vector space.} is parametrized by the fixed laser parameters 
$\theta$ and $\varphi$, as captured by the unit vector ${\bf n} = (\sin \theta \cos \varphi, 
\sin \theta \sin \varphi, \cos \theta)$. Since the evolution is purely geometric, the holonomic 
one-qubit gate $U(C_{\bf n})$ associated with $C_{\bf n}$ coincides with the action of the 
time evolution operator $U (\tau,0)$ on $\mathcal{M}$, i.e., $P_{\mathcal{M}} 
U(\tau,0) P_{\mathcal{M}}$ with $P_{\mathcal{M}} = \ket{0} \bra{0} + 
\ket{1} \bra{1}$. Explicitly, we find \cite{sjoqvist12}
\begin{eqnarray}
U (C_{\bf n}) = \ket{d} \bra{d} - \ket{b} \bra{b}  
\label{eq:idealgateoperator}
\end{eqnarray} 
with the dark and bright states $\ket{d} = \cos \frac{\theta}{2} \ket{0} + e^{-i\varphi} 
\sin \frac{\theta}{2} \ket{1}$ and $\ket{b} = e^{i\varphi} \sin \frac{\theta}{2} \ket{0} - 
\cos \frac{\theta}{2} \ket{1}$, respectively, which constitute another orthonormal frame 
spanning $\mathcal{M}$. We may write 
\begin{eqnarray}
U(C_{\bf n}) = i e^{- i\frac{1}{2} \pi {\bf n} \cdot \boldsymbol{\sigma}} = 
{\bf n} \cdot \boldsymbol{\sigma} 
\label{eq:idealgatematrix}
\end{eqnarray} 
with $\boldsymbol{\sigma} = (\sigma_x,\sigma_y,\sigma_z)$ the Pauli operators 
$\sigma_x = \ket{0} \bra{1} + \ket{1} \bra{0}$, $\sigma_y = -i \ket{0} \bra{1} + 
i\ket{1} \bra{0}$, and $\sigma_z = \ket{0} \bra{0} - \ket{1} \bra{1}$. Here, $U(C_{\bf n})$ 
is the nonadiabatic non-Abelian geometric phase \cite{anandan88} associated with the 
path $C_{\bf n}$. 

$U(C_{\bf n})$ corresponds to a $\pi$ rotation of the qubit around the direction ${\bf n}$. 
To obtain an arbitrary SU(2) operation, this gate must therefore be combined with another 
holonomic gate produced by a second pulse pair. To see this, assume that the laser 
parameters of two sequentially applied pulse pairs define unit vectors ${\bf n}$ and 
${\bf m}$. The combined gate reads 
\begin{eqnarray}
U(C_{\bf m}) U(C_{\bf n}) = {\bf m} \cdot {\bf n} - i \boldsymbol{\sigma} \cdot 
({\bf n} \times {\bf m}) . 
\end{eqnarray}
This corresponds to a rotation angle $2\arccos \left( {\bf n} \cdot {\bf m} \right)$ 
around the rotation axis ${\bf n} \times {\bf m}$, i.e., an arbitrary SU(2).  

Let us now turn to the off-resonant case where $\delta \neq 0$. Here, the geometric nature 
of the evolution of $\mathcal{M}$ depends on the detailed form of the pulse envelope $F(t)$ 
since $H(t)$ may no longer commute with the time evolution operator\footnote{The geometric 
nature depends on the pulse form since the extra term $\delta \ket{e} \bra{e}$ in the Hamiltonian 
$H(t)$ makes it in general necessary to use time ordering to compute the time evolution 
operator. As a consequence, $H(t)$ does not necessarily commute with the associated time 
evolution operator, which in general implies that $U^{\dagger} (t,0) H(t) U(t,0)$  would 
not vanish on $\mathcal{M}$ and thereby creating a nontrivial non-Abelian dynamical phase 
acting on the qubit. The square pulse takes care of this since the time evolution operator acts 
trivially on $\mathcal{M}$ before and after the pulse and can be computed without time 
ordering during the pulse.}. However, there is one physically justified choice where the 
evolution of $\mathcal{M}$ is purely geometric, viz., when $F (t)$ is a square pulse, i.e., 
$F (t) = F_0$ for $0 \leq t \leq \tau$ and zero otherwise. For such a pulse, consider evolution 
between $t_0$ and $t_1$, where $t_0 \leq 0$ and $t_1 \geq \tau$. The corresponding 
time evolution operator reads: 
\begin{eqnarray}
U(t_1,t_0) & = & U(t_1,\tau) U(\tau,0) U(0,t_0) = 
e^{-i(t_1-\tau)\delta \ket{e} \bra{e}} U(\tau,0) e^{it_0 \delta \ket{e} \bra{e}}  
\end{eqnarray}
with 
\begin{eqnarray}
U(\tau,0) = e^{-i\tau (F_0 H_0 + \delta \ket{e} \bra{e})} . 
\end{eqnarray}
The action of $U(t_1,t_0)$ is trivial on $\mathcal{M}$ on $t_0 < t < 0$ and $\tau < t < t_1$, 
provided $\tau$ is chosen such that $\mathcal{M}$ undergoes cyclic evolution. Thus, it is 
sufficient to consider $U(\tau,0)$ in the following.  

Since $\bra{k} U^{\dagger} (t,0) \left( F_0 H_0 + \delta \ket{e} \bra{e} \right) 
U(t,0) \ket{l} = \bra{k} \left( F_0 H_0 + \delta \ket{e} \bra{e} \right) \ket{l} = 0$, $k,l=0,1$, 
on $0 \leq t\leq \tau$, the nontrivial part $U(\tau,0)$ of the time evolution operator is purely 
geometric on the single-qubit subspace $\mathcal{M}$. It further corresponds to cyclic 
evolution with period $\tau = 2\pi / \sqrt{\delta^2 + 4F_0^2}$ for which we find holonomic 
gate 
\begin{eqnarray}
U({\bf n},\chi) = \ket{d} \bra{d} - e^{-i\chi}  \ket{b} \bra{b} = 
e^{i\frac{1}{2} (\pi - \chi)}  e^{-i\frac{1}{2} (\pi - \chi) {\bf n} \cdot \boldsymbol{\sigma}}
\label{eq:nonidealgateoperator}
\end{eqnarray}
with 
\begin{eqnarray}
\chi = \frac{\pi \delta}{\sqrt{\delta^2 + 4F_0^2}} . 
\label{eq:chi}
\end{eqnarray} 
Up to the unimportant overall phase factor $e^{i\frac{1}{2} (\pi - \chi)}$, we 
see that $U({\bf n},\chi)$ corresponds to a single-qubit rotation with angle $\pi - \chi$ 
around the direction ${\bf n}$. Thus, an arbitrary single-qubit operation can be reached 
by independently varying the detuning and laser parameters $\theta$ and $\varphi$. In 
particular, $U({\bf n},\chi)$ connects to the identity in the $\delta /(2F_0) \rightarrow \infty$ 
limit\footnote{In the small rotation angle limit, $\delta$ should thus be very large compared 
to the pulse strength $F_0$. By comparing with Eq. (\ref{eq:chi}), we see that the duration 
$\tau$ becomes very short in this limit, which may invalidate the rotating wave approximation. 
This in turn implies that the geometric nature as well as the stability of the gate would be 
lost \cite{spiegelberg13}. Thus, gates corresponding to small qubit rotations are probably 
difficult to realize in practise by using our scheme.}, i.e., $U({\bf n},\chi \rightarrow \pi) = 
\hat{1}$, and it coincides with the nonadiabatic holonomic gate proposed in Ref.~\cite{sjoqvist12} 
in the resonant case, i.e., $U({\bf n},0) = U(C_{\bf n})$.

Further insight into the geometry of $U({\bf n},\chi)$ can be obtained by calculating 
the connection 1-form associated with the time-evolved computational subspace, i.e., 
$\mathcal{M} (t) = \textrm{Span} \{ U(t,0) \ket{d}, U(t,0) \ket{b} \}$ with $\mathcal{M} (\tau) = 
\mathcal{M} (0) = \mathcal{M}$. By solving the Schr\"odinger equation, we find 
\begin{eqnarray} 
\ket{d(t)} & = & U(t,0) \ket{d} = \ket{d} , 
\nonumber \\ 
\ket{b(t)} & = & U(t,0) \ket{b} 
 \nonumber \\ 
 & = & e^{-i\frac{1}{2}\delta t} \left( e^{-i\frac{1}{2}\sqrt{\delta^2 + 4F_0^2}t} 
\sin \nu \ket{+} + e^{i\frac{1}{2}\sqrt{\delta^2 + 4F_0^2}t} \cos \nu \ket{-} \right) , 
\end{eqnarray}
where 
\begin{eqnarray} 
\tan \nu = \frac{\delta + \sqrt{\delta^2 + 4F_0^2}}{2F_0} 
\label{eq:nu}
\end{eqnarray}
and $\ket{\pm}$ are the bright eigenstates of $H_0 + \delta \ket{e} \bra{e}$. Clearly, 
the only nonzero component of the vector potential 
\begin{eqnarray}
A(t) = i\sum_{k,l=d,b} \bra{k(t)} \dot{l}(t) \rangle \ket{k(t)} \bra{l(t)} = 
\sum_{k,l=d,b} A_{kl} (t) \rangle \ket{k(t)} \bra{l(t)} 
\end{eqnarray} 
is $A_{bb} (t)$. We find, 
\begin{eqnarray} 
A_{bb} (t) = - \sqrt{\delta^2 + 4F_0^2} \sin^2 \nu . 
\end{eqnarray}
The holonomy can be obtained as $\ket{d} \bra{d} + e^{i\gamma} \ket{b} \bra{b}$, 
where $\gamma$ is the Aharonov-Anandan geometric phase \cite{aharonov87} associated 
with $\ket{b(t)}$, i.e., 
\begin{eqnarray} 
\gamma & = & \int_0^{\tau} A_{bb} (t) dt = -2\pi \sin^2 \nu = 
-2\pi \frac{1}{2} \left( 1 + \frac{\delta}{\sqrt{\delta^2 + 4F_0^2}} \right) 
\nonumber \\ 
 & = & \pi - \chi , \ \textrm{mod} (2\pi) ,  
\end{eqnarray}
where we have used Eq. (\ref{eq:nu}). Thus, $\ket{d} \bra{d} + e^{i\gamma} \ket{b} \bra{b}$ 
coincides with $U({\bf n},\chi)$. 

Note that $A(t)$ commutes with itself within each pulse pair. This is the underlying reason 
why the holonomy $U({\bf n},\chi)$ can be understood in terms of the Abelian geometric 
phase factor $e^{i\gamma}$ of $\ket{b(t)}$. However, $[A(t),\tilde{A}(t')]$ can be nonvanishing 
for $A$ and $\tilde{A}$ evaluated for two subsequent pulse pairs with different laser 
parameters. To see this, consider $A(t) = A_{bb} (t) \ket{b(t)} \bra{b(t)}$ and $\tilde{A}(t) = 
A_{\tilde{b}\tilde{b}} (t) \ket{\tilde{b} (t)} \bra{\tilde{b} (t)}$ with $\ket{b(t)}$ and 
$\ket{\tilde{b} (t)}$ corresponding to two different sets of laser parameters 
$\theta,\varphi,\delta$ and $\tilde{\theta},\tilde{\varphi},\tilde{\delta}$, respectively. 
We obtain
\begin{eqnarray} 
[A(t),\tilde{A}(t')] & = & A_{bb} (t) A_{\tilde{b}\tilde{b}} (t') 
\nonumber \\ 
 & & \times \left( \ket{b(t)} \bra{b(t)} \tilde{b}(t) \rangle 
\langle \tilde{b}(t)| - |\tilde{b}(t) \rangle \langle \tilde{b}(t) \ket{b(t)} \bra{b(t)} \right) \neq 0 , 
\end{eqnarray}
where $t$ and $t'$ belong to the support of the respective two laser pulse pairs.  
This proves the non-Abelian nature of the gate. 

Before concluding, let us briefly comment on how the restriction to square-shaped 
pulses may influence the flexibility of our scheme. The key point here is that it should 
be possible to vary freely the detuning, amplitude, and phase of the pulses, despite this 
restriction. To see explicitly what this means, let us consider a possible implementation of 
our scheme in which transitions between two atomic levels $j=0,1$ and an excited state 
$e$ are induced by pulsed electric fields ${\bf E}_j (t) = g_j (t) \cos (\omega_j t) \boldsymbol{\epsilon}_j$, 
where $\boldsymbol{\epsilon}_j$ are the polarizations. Here, the time dependent 
part consists of two factors: $g_j(t)$ determining the pulse shape and $\cos (\omega_j t)$ 
determining the detuning $\delta_j = \omega_{je} - \omega_j$, $\omega_{je}$ being 
the energy difference (in units where $\hbar = 1$) between the bare energy eigenstates 
$\ket{e}$ and $\ket{j}$. The scheme requires that $g_0 (t) = g_1 (t) \propto F(t)$ being 
square-shaped and $\delta_0 = \delta_1 = \delta$. The amplitude and phase parameters 
$\theta$ and $\varphi$ are determined by the ratio $\bra{e} \boldsymbol{\mu} 
\cdot \boldsymbol{\epsilon}_0 \ket{0} / \bra{e} \boldsymbol{\mu} \cdot 
\boldsymbol{\epsilon}_1 \ket{1} = - e^{-i \varphi} \tan \frac{\theta}{2}$, $\boldsymbol{\mu}$ 
being the electric dipole operator. We thus see that the required flexibility is obtained in this 
particular setting if the polarization $\boldsymbol{\epsilon}_j$ and oscillation frequency 
$\omega_j$ of each electric field pulse can be varied independently, although $g_j (t)$ is 
restricted to having square shape. 

In conclusion, we have demonstrated holonomic single-qubit gates in off-resonant 
$\Lambda$ system. These gates can be used to implement any single-qubit gate and 
would, together with an entangling holonomic two-qubit gate, constitute an all-geometric 
universal set of gates. The off-resonant holonomic gates require square-shaped pulses 
in order to preserve the purely geometric nature. Our finding implies that the assumption 
of zero detuning in the original scheme \cite{sjoqvist12} is not a necessary requirement 
to perform nonadiabatic holonomic quantum computation in $\Lambda$ systems. 
We further note that the additional flexibility associated with the detuning makes it 
possible to perform arbitrary single-qubit operations by a single pulse pair. This latter 
feature may help experimental realizations of holonomic quantum computation as it  
reduces the number of pulses needed to implement arbitrary single-qubit operations. 
The scheme can be implemented experimentally in various systems, such as trapped 
atoms or ions, superconducting qubits, or NV-centers in diamond. 
\vskip 0.3 cm 
Financial support from the Swedish Research Council (VR) through Grant No. D0413201 
is acknowledged. 

\end{document}